\documentclass[a4paper]{jpconf}
\usepackage{amssymb}
\usepackage{graphicx}
\usepackage{amsthm}
\usepackage{latexsym}
\newcommand\beq{\begin{equation}}
\newcommand\eeq{\end{equation}}
\newcommand\bea{\begin{eqnarray}}
\newcommand\eea {\end{eqnarray}}

\newcommand{\ie}{\emph{i.e.},\,}

\def\un\a{{\underline\alpha}}

\def\a{{\alpha}}

\def\e{{\epsilon}}

\begin{document}
\title{\textbf{Causal Sets: Quantum gravity from \\a fundamentally discrete spacetime
}}

\author{\textbf{Petros Wallden}}
\address{1. University of Athens, Physics Department, Nuclear \& Particle Physics Section, Panepistimiopolis 157-71, Ilissia
Athens, Greece}

\address{2. Chalkis Institute of Technology, Department of Technological Applications,\\  Psahna-34400, Greece}

\ead{\textbf{petros.wallden@gmail.com}}

\date{}

 \vspace{1cm}

\begin{abstract}

In order to construct a quantum theory of gravity, we may have to
abandon certain assumptions we were making. In particular, the
concept of spacetime as a continuum substratum is questioned. Causal
Sets is an attempt to construct a quantum theory of gravity starting
with a fundamentally discrete spacetime. In this contribution we
review the whole approach, focusing on some recent developments in
the kinematics and dynamics of the approach.
\end{abstract}

\section{Motivation}

The challenge to construct a quantum theory of gravity is great.
There exist many approaches, each of them having their merits,
however there is no widely accepted and confirmed theory. The
difficulty of the problem suggests, similarly with other paradigms
in the history of physics, that we may need to abandon some of our
a-priori assumptions about reality. In particular the continuity of
spacetime that we assume, is questioned in several places.
Approaches to quantum gravity that start from a continuum theory,
observe a (possibly effective) discreteness\footnote{Discreteness of
volume in Loop Quantum Gravity, dualities between very small and
large scales in string theory, etc.}. Another field that we may get
hints about quantum gravity, is the black holes thermodynamics.
Infinities that arise in this case, could be avoided if space was
discrete as it is observed in\cite{entanglement
entropy}\footnote{Research in causal sets, involves some treatment
of black hole thermodynamics. In this contribution we will not deal
with it, but we give the following references for the interested
reader \cite{BH causet}.}. The infinities we face in quantum field
theory (though we have learnt to live with them), and singularities
in general relativity are two other reasons why discreteness could
simplify our physical picture. All the above, would be resolved, if
one was willing to abandon spacetime continuity\footnote{There are
also models for modifying gravity that would resolve the
cosmological constant problem, that favor discreteness.}. It is
therefore a natural starting point, for building a new quantum
theory of gravity, to take spacetime fundamentally discrete.

The second element in causal sets, is \emph{causality}. Here again,
it seems more intuitive to speak of causal relations (that have the
natural interpretation of cause and effect), rather than some
standard spacetime where one needs to assume topology, differential
structure and metric. After all, as we will see below causal
relations encode most of the information of the metric.

We therefore, come to conclude that it may be better starting a
quantum gravity theory from a discrete spacetime that comes along
with the causal relations of the elements (points) of this
spacetime. This is precisely the starting point of the causal set
approach to quantum gravity. It was originally introduced in
\cite{blms}.

\section{Definition}

Mathematically a causal set (or causet) is a set $C$ endowed with a
partial order relation $\prec$ which is irreflexive ($x \nprec x$),
transitive ($x \prec y \prec z \Rightarrow x \prec z$), and locally
finite $[x,y] \equiv \left(|\{y | x \prec y \prec z \}| < \infty \;
\forall\quad x, z \in C\right)$ (where $|A|$ indicates cardinality
of the set $A$).  The local finiteness condition imposes the
requirement of discreteness, because it requires that there is only
a finite number of elements between every pair in the causet. The
partial order relation, represent the causal relation between
elements of $C$, so if $x\prec y$ it means that $x$ is at the past
of $y$.

The next thing to do is to see why a causal set (as defined above)
could possibly replace continuum spacetime. This is based on a
theorem by Malament \cite{Malament}:

\begin{quote}The metric of a globally hyperbolic spacetime can be
reconstructed uniquely from its causal relations up to a conformal
factor.
\end{quote}
The theorem suggests that the causal structure encodes most of the
information of the metric. What we lack to obtain the full metric is
a way to fix the conformal factor, in other words to fix the volume.
This is precisely done by considering a discrete spacetime. In
causal set we make the correspondence $\textrm{Volume= Number of
elements}$\footnote{In particular each element of the causal set
corresponds to 1 Plank unit Volume.}. By having only a finite number
of causal set elements between any pair of causally related
spacetime points, we can use that number to fix the volume. We thus
have the ``slogan'' Order+Number=Geometry. We are now in position to
make the central conjecture of causal sets (also called the
Hauptvermutung):

\begin{quote}\textbf{Central Conjecture}: Two distinct, non-isometric
spacetimes cannot arise from a single causal set.\end{quote}

The conjecture, while it still remains a conjecture has very solid
theoretical and numerical evidence in its support\footnote{Most of
the work on the kinematics of causal sets, regarding the dimension
(e.g. \cite{dimension}), timelike and spacelike distances
(\cite{bg,diameters}), topology (\cite{homology}), provide support
to the conjecture.}. The central conjecture guarantees that there
can be only one (essentially) spacetime approximating a given causal
set. Here we must stress the fact that a causal set behaves as a
continuum manifold, under several conditions, and in particular,
when the number of causal set elements is very big. In this case we
speak of the \emph{continuum approximation}\footnote{It is the
analogue of continuum limit of other approaches, however it is not
really a limit, since we never take the discreteness scale going to
zero, as it is done for example in CDT \cite{CDT}.}.

We define the concept of \emph{faithful embedding} to make precise
when a continuum spacetime approximates a causal set:

\begin{quote} A faithful embedding is a map $\phi$ from a causal set
$\mathcal P$ to a spacetime $\mathcal M$ that:\begin{enumerate}

\item preserves the causal relation (\ie $x\prec y\iff
\phi(x)\prec \phi(y)$) and
\item is ``volume preserving'', meaning
that the number of elements mapped to every spacetime region is
Poisson distributed, with mean the volume of the spacetime region in
fundamental units, and
\item $\mathcal M$ does not possess curvature
at scales smaller than that defined by the ``intermolecular
spacing'' of the embedding (discreteness scale).\end{enumerate}
\end{quote}
The central conjecture therefore reads as ``a causal set cannot be
faithfully embedded in two non isometric spacetimes''. Taking a
closer look at the definition of faithful embedding, we see that a
regular lattice cannot correspond to a faithful embedding of
Minkowski spacetime. It fails because, in a regular lattice,
considering a very boosted frame, we will end up with big volumes
having no elements at all and fail to satisfy the condition (ii) of
faithful embedding. Instead, a \emph{random} lattice would do this
job. For example one generated by sprinkling elements in spacetime
randomly with probability $P(n)=\frac{(\rho v)^n\exp^{-\rho
V}}{n!}$, in other words a Poisson sprinkling\footnote{And the
expected variance is of course $\sqrt V$.}.

Causal sets are constructed in such a way, that they respect Lorentz
invariance at the kinematic level already \footnote{Unlike for
example spin networks, where Lorentz invariance is expected to arise
due to some properties of the superimposed Lorentz-violating spin
networks}. These two observations are analyzed in detail in
\cite{lorentz_invariance}.

\section{Kinematics}

In all discrete/combinatoric approaches to quantum gravity one
important step is the so called ``inverse problem''
\cite{inverse_problem}. It consists of two parts, the first is how
do continuum like structure arise from the fundamentally discrete
base. The second part, is how shall we recognize that we do have
something continuum like (and what) when we actually have it.

In our case, we want to address this issue for a causal set. In
particular we want to obtain continuum notions from the partial
order. A typical causal set (if taken by pure counting and
attributing equal weights to all causal sets), it is a
Kleitman-Rothschild order \cite{KR order}. These orders have only
three layers (i.e. longest chain is a 3-element chain) and 1/4 of
elements sit at the first layer, 1/2 in the second and 1/4 in the
third. This is certainly not manifold-like (it contains only
3-moments of time). We thus need the dynamics to select for us a
causal set that looks like a continuum manifold. For now we will
deal with properties of causal sets that in the case there exist a
manifold that faithfully embeds to it, correspond to continuum
properties such as lengths of curves.

Some definitions are in order: \\ \textbf{Definition 1:} A pair of
elements $x,y$ is a \emph{link} if $x\prec y$ and $\nexists\quad
z|x\prec z\prec y$. In other words is the most basic
relation, that cannot be implied by transitivity.\\
\textbf{Definition 2:} Chain $C$ is a collection of elements such
that $\forall x,y\in C$  either $x\prec y$ or $y\prec x$. In other
words it is linearly ordered \footnote{this intuitively corresponds
to a timelike curve at the continuum}.\\ \textbf{Definition 3:} We define a \emph{path} to be a chain, where each pair of consecutive elements are links.  \\
\textbf{Definition 4:} We define a \emph{2-link}, given a pair of
unrelated elements $x,y$, an element $z$ that $x,z$ and $y,z$ are
both links.

The concept of timelike curve is represented by a chain. Moreover,
it can be shown that we can define the timelike distance between two
related elements $x\prec y$ of the causal set. In the continuum it
corresponds to the maximum length of timelike curve that joins $x$
and $y$. In the causal set this is simply the chain of maximum
cardinality starting from $x$ and ending at $y$
\cite{bg}\footnote{To be more precise it is proportional and the
proportionality constant in general depends on the dimension}.

\beq d_t(x,y)=\max|C|  \textrm{ where }C \textrm{ start at x and
ends at y}\eeq It has been tested theoretically but also numerically
\cite{bg,diameters}, that in the case that the causal set is
faithfully embedded at a manifold \footnote{Basically, it was tested
that causal sets that arose from sprinkling to some continuum
manifold} the above distance agrees with the continuum spacelike
distance, in the limit of a sufficiently big causal set.

To get spacelike distance for a causal set that corresponds to
Minkowski spacetime is considerably more difficult. Lee Smolin for
example has claimed that what lacks from causal sets is some spatial
information. The first naive attempt to get the spatial distance
would be the following. Consider $x,y$ two unrelated (spacelike)
elements of the causal set. Then let $u\in J^+(x)\cap J^+(y)$ and
$v\in J^-(x)\cap J^-(y)$. \beq d_s(x,y)=\min d_t(v,u)\eeq
Unfortunately this fails as was already noted in \cite{bg} and
confirmed numerically in \cite{diameters}. A more refined treatment
is required where $u\in J^+(x)\cap J^+(y)$ but also $u$ is a 2-link
(w.r.t. $x,y$). The reader is referred too \cite{diameters} for
further details.

It can be also shown, that using this notion of spacelike distance,
one can define closest spatial neighbors (timelike closest neighbors
are simply the links). Using this concept we can measure the length
of any curve in the causal set (not only on chains). More
importantly this generalizes in curved spacetimes, provided that
there is no curvature at very small scales. This is a reasonable
assumption made also by Gibbons in \cite{Gibbons}.

The dimension of a causal set is also examined in \cite{dimension}.
Assume we have two related elements $x\prec y$, with $d_t(x,y)=n$
and the number of elements between them being N ($|V_{xy}|=N\textrm{
where } z\in V_{xy} \textrm{ if } x\prec z\prec y$). We then see how
the volume scale as we double the timelike distance \footnote{in
other words take an element $z$ such that $d_t(x,z)=2n$ and find
$|V_{xz}|$ }. This and other dimension measures that work for causal
sets embedable in continuum spacetimes as well as extensions of the
notion of dimension for non-manifold like causal sets can be found
in \cite{dimension}.

Finally, the spatial topology has been analyzed (e.g. in
\cite{homology}). The concept of spatial slice is replaced by the
following: We consider a \emph{maximal anti-chain} a collection of
elements $\{e_i\}$ such that $e_i\nprec e_j$ and does not exist any
other element being unrelated to all the elements of our
anti-chain\footnote{Adding any other element makes our set stop
being an anti-chain}. However, this is not a good analogue for a
spatial slice. We need to ``thicken'' by considering all elements
with distance a number $n$ from the anti-chain. There is a nice way
to recover the topology of this slice, using simplicial compleces.
The results are in very good agreement with continuum results for
causal sets that embed in continuum spacetimes \cite{homology}.

\section{Dynamics}
While a lot can be said from the kinematic sector of a theory (and
as we will see later, some phenomenological effects), the true
physical content of a theory lies in the dynamics. There are two
ways to try and adopt the dynamics. The first\footnote{called
initially by Sorkin as the ``principled'' approach} takes as
fundamental ingredient the causal set itself and attempts to
generate the full causal set from some basic principles that are in
some sense natural to the causal set approach, in other words simply
related with the partial order. The second approach \footnote{called
the ``opportunistic'' approach}, is an attempt to guess some
dynamics (possibly by rephrasing the continuum Einstein action, in
terms of the order relation) and get ``correct'' result\footnote{It
is opportunistic, since the Einstein action has a very good
motivation in terms of continuum spacetime, but the analogue on a
causal set is not something intuitive.}. Finally, of interest is to
consider, quantum matter (or fields) on one classical causal set (an
analogue of QFT in curved spacetime). There has been considerable
progress in the latter recently. Here we will first consider some
classical (alas stochastic dynamics). Then we will discuss about
quantum dynamics. This task has not been fully accomplished, and we
will only mention some toy models, attempts and directions for
future research. Finally we will deal with quantum matter on a
causal set.

\subsection{Classical Dynamics for Causal Sets}
\begin{figure}
\scalebox{0.40}{\includegraphics{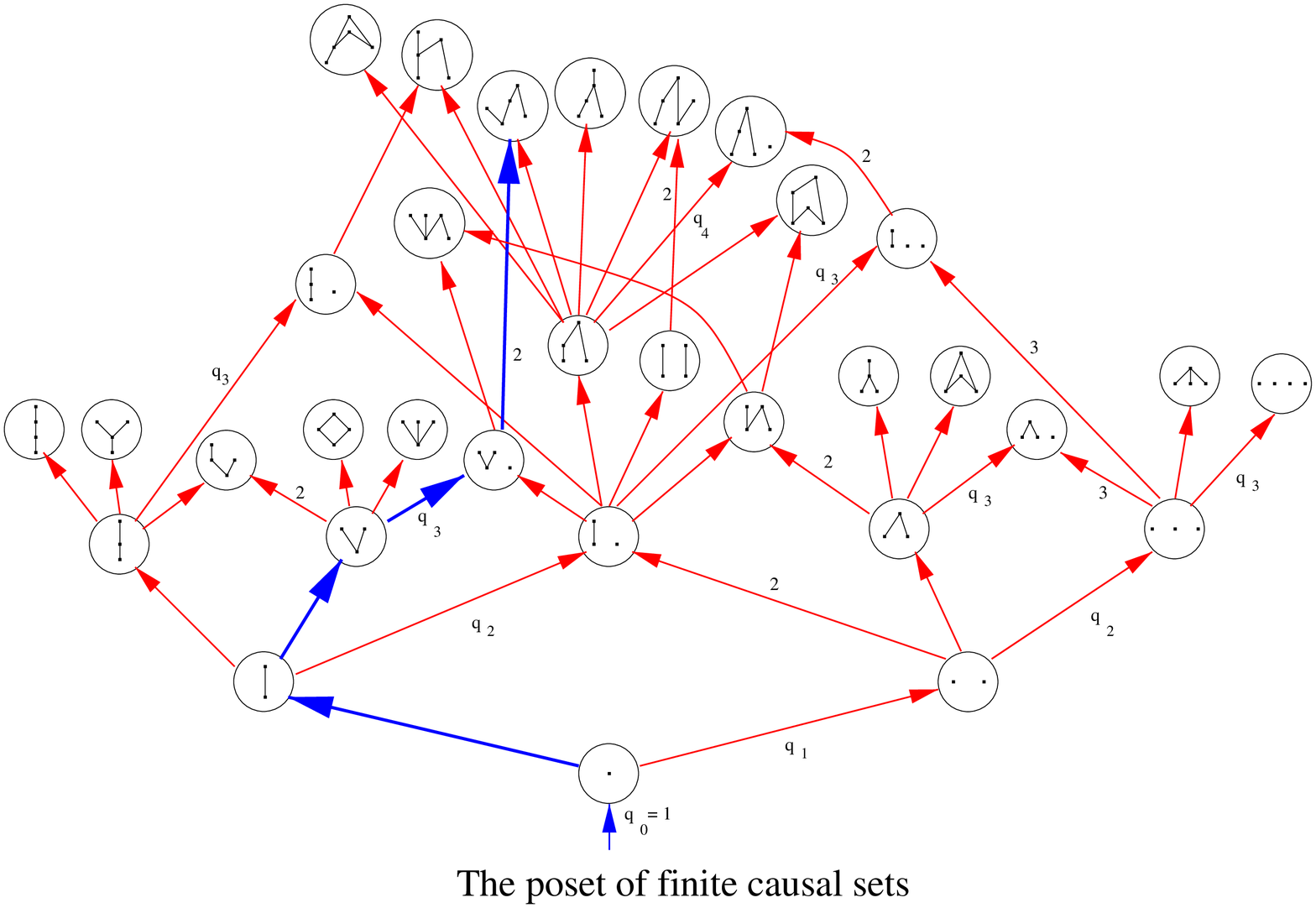}} \label{poscau}
\end{figure}

The stochastic classical dynamics have been analyzed in depth (see
\cite{CGD}). One can construct a partial order that consists of all
the causal sets (a partial order of partial orders), where two
elements are related if the one is subset (when viewed as a partial
order) of the other. It is called poscau (see Fig. \ref{poscau}).
This is the arena for causal set dynamics and is an analogue of the
superspace. It is the set of all possible spacetimes. We can imagine
a process that generates all the causal sets, by growing the causal
set element by element. This process is characterized by some
probability, that each new-born element will be at the future or
spacelike (unrelated) from each of the already existing elements.
These processes are know as \emph{(classical) growth dynamics}
(CGD)\footnote{They can be thought as generalizations of random
walks}. We impose some physical conditions that these transitions
probabilities must obey. In particular the requirements are:

\begin{enumerate}
\item \textbf{Internal temporality}. Each new element is born to the future or spacelike to all
existing elements (never at the past).
\item \textbf{Discrete general covariance}. The probability of arriving at a
particular causal set does not depend on the order the elements are
born. In other words the probability does not depend on the path
chosen on the Figure.
\item \textbf{``Bell's causality''}. This condition states that the probability of new-born
elements depends on the past of the new-born element and thus is not
affected by births in spacelike regions. The name cames from the
fact that it is related with the paradoxes that arise in quantum
theory due to non-locality. It is believed that this condition may
need to be relaxed (or transformed) if we were to consider a quantum
theory of causal sets\footnote{However, it is not necessary since
non-local effects on a causal set, may arise even if the dynamics of
the causal set itself are in some sense local.}.
\item \textbf{The Markov Sum rule}. This means that the sum of all
probabilities starting from a particular causal set (element of the
poscau), sum up to one.
\end{enumerate}

These conditions turn out to be very restrictive, and there is only
one free parameter left for each level\footnote{By level, we mean
the number of total elements of the causal set.}. One particular
choice of these parameters results in the transitive percolation
that has been studied in depth (see references in \cite{CGD}). The
different choices give rise to very different causal sets. While
some of these have striking similarities with cosmological models
\cite{causet De-Sitter}, the resulting causal sets are not in
general manifold like and thus we would need to resort to quantum
dynamics to explain the continuum like behavior we see.

\subsection{Quantum Causal Sets}

A first attempt to quantize the above dynamics, would be to use
quantum amplitudes instead of probabilities. In other words to allow
the transition probabilities to be complex
numbers\footnote{Tentative results show that the typical causal set
still is not one manifold-like}.This would be a generalization of
what is known as quantum random walk (QRW). The whole causal sets
program is based on spacetime rather than space. It is not possible
to do a canonical quantization, since there is no well defined way
to break up the full causal set to spatial slices with an external
time parameter\footnote{Despite the fact that as we have seen there
exist an analogue of a spacelike surface for causal sets that
correspond to spacetimes that allow such a foliation. However is not
fundamental for the approach.}. It is generally believed that in
order to deal with quantum causal sets, one is forced to resort to
sum over histories quantization. The mathematical ingredient one
needs to use is the \emph{quantum measure} \cite{quantum
measure}\footnote{In classical stochastic theories such as the
Brownian system, one uses again histories, and a measure (the Wiener
measure) on the set of possible histories. In a sense we believe
that quantum theory comes as a natural generalization of classical
stochastic theories rather than deterministic theories.}. To
interpret it properly, one needs to address the basic problems of
the foundations of quantum theory. Consistent (or decoherent)
histories is one approach, while there is considerable recent
activity in a novel interpretation the ``Pimbino'' interpretation
(also called the co-event or anhomomorphic logic interpretation)
\cite{Piombino interpretation}.

The most interesting result however comes from considering
2-dimensional partial orders \cite{Brightwell:2007}. Here the
dimension is not to be understood as spacetime dimension, it is
rather a technical term for partial orders. In particular it means
the following: Assume that we have a set $\mathcal
P=\{e_1,e_2,\cdots\}$ and we have two linear orderings of this set .
Then we take the intersection of this linear orderings, in other
words an element $e_1\prec\e_2$ iff $e_1$ is before $e_2$ in both
linear order we have. The resulting partial order is called
2-dimensional, if it can be generated from the intersection of 2
linear orderings\footnote{By analogy it is n-dimensional if it is
the intersection of n linear orderings.}. For 2-dimensional partial
orders it turns out, that they do correspond to 2-dimensional
spacetimes, however this analogy breaks down for higher dimensions.

Brightwell et al in \cite{Brightwell:2007} considered 2-d orders and
atttempted to find what the typical 2-d order would look like. Note,
that as we have already mentioned a typical causal set is not
manifold-like (it is a Kleitman-Rothschild order). This problem is
also known as the entropy problem and is faced in other approaches
such as the DT (dynamical triangulation) and a possible solution is
to restrict to sum over sum particular class rather than all
possible configurations (as was done in CDT (causal dynamical
triangulations) \cite{CDT}). Interestingly, in 2-d orders the major
contribution comes from causal sets that are manifold-like and
moreover there are faithfully embeddable in 2-d Minkowski. This can
be seen as a toy model to see how manifold-like behavior (and indeed
flat!) arises from simple considerations on a causal set.

Finally, recently Bombelli and Svedrlov in \cite{sverdlov} have
taken the ``opportunist'' approach. They have written the Einstein
action in terms of causal set elements, and have used this weight to
make a sum over all possible causal sets. However, this is work is
at the beginning. One has to see what is a typical causal set, if we
were to use this weight and also what consequences that would have.
The reader is referred to the original papers for further details.

\subsection{Quantum Matter on Causal Set}

One thing of great interest, would be to see how matter and fields
behave, on a single fixed (and thus classical) causal set. This
question corresponds to fixing the background to some (generally)
curved spacetime and allowing quantum fields on it. For causal sets,
we need to firstly see how we recover standard results, and then see
wether the assumption of discrete background may have some effect
that could be possible to test and thus confirm or falsify the
causal set assumption\footnote{Of course the observed effect,
typically would depend on some parameters and perhaps on the full
quantum dynamics.}.

The first attempt would be to consider point (massive) particles
moving along a chain of the causal set that faithfully embeds at
Minkowski spacetime. There are several models considered
\cite{DHS,DPS}, that in all of those, when taking the continuum
approximation, the particle follows approximately timelike geodesic,
but deviating (swerving) slightly. In other words it is like having
some drift, and all models result in a diffusion equation depending
on a single parameter the diffusion strength $k$. Let us here
briefly describe the first model (\cite{DHS}). The particles
trajectory is a chain $\{e_1,e_2,\cdots\}$ in the causal set. we
define a forgetting time $t_f$, which is the time below which the
causal set may behave non-locally, but above which it behaves
normally. If the particle has position $e_n$ with four momentum
$p_n$ the next element is chosen

\begin{itemize}
\item $e_{n+1}$ is at the causal future of $e_n$ and at proper time
$t_f$
\item the momentum change $|p_n-p_{n+1}|$ is minimized (where $p_{n+1}$ is on the mass shell and proportional to the vector between $e_n$ and $e_{n+1}$).
\end{itemize}

The resulting motion, heuristically, stays as close to straight as
possible, having however some random fluctuations in the particles
momentum. The diffusion equation resulting in the continuum
approximation was (see \cite{DHS})

\beq \frac{\partial\rho}{\partial\tau}=k\nabla^2\rho-\frac1m
p^\mu\partial_\mu\rho \eeq

Following similar lines, one may attempt to see what happens for
massless particles. The added difficulty being that the trajectory
is no longer a chain (which by definition is a timelike curve). The
resulting equation has two parameters (diffusion and drift
parameters) $k_1, k_2$ and there is again some non-locality scale.
The reader is referred to \cite{DPS}.

This particular feature, of some non-locality scale, appears also
from consideration of particles propagators. In order to do this one
needs to define the D'Alembertian operator $\square$. In
\cite{Sorkin:dalembertian}, the d'Alembertian is introduced. We will
follow briefly \cite{johnston}, to obtain the retarded propagator
for a particle on a causal set.

To get the propagator $K(x,y)$ from an element $x$ to $y$ in its
future, we need to sum over all the chains (or paths) on a causal
set from $x$ to $y$ with a particular amplitude (c.f. quantum
mechanics where one adds the quantum amplitudes to go from one point
to another and sums over all the possible trajectories). We need to
make a choice of which trajectories to sum over (either all chains
or all paths), and what weight to put on each trajectory. These
choices have to be made in such a way, that when a causal set is
faithfully embedable in Minkoweski spacetime, to give us back a
propagator of the Klein-Gordon equation (in particular, with the
choices made we shall get the retarded propagator). In particular
the amplitude depends on two parameters $a$ being the probability
that the particle would `hop' once along the trajectory from one
element to another, and $b$ being the probability that would stop at
an element of the trajectory. For a chain of length $n$ (where we
have $n$ hopes and $n-1$ intermediate stops) the amplitude would be
$a^nb^{n-1}$. For causal set in 1+1 Minkowski and 3+1 Minkowski we
fix the values of $a$ and $b$ to be respectively\footnote{The values
depend only on the mass of the particle and the volume corrseponding
to each causal set element, i.e. the density.}:

\beq a=\frac12,\quad b=-\frac{m^2}\rho\eeq

\beq a=\frac{\sqrt\rho}{2\pi\sqrt6},\quad b=-\frac{m^2}\rho \eeq
With these choices the retarded Klein-Gordon propagator is obtained
as shown in \cite{johnston}.

Building up on this work, using the retarded Klein-Gordon
propagator, Johnston in \cite{johnston2} proceeded and considered
scalar quantum field on a causal set, and he computed the Feynman
propagator\footnote{Of crucial importance was the use of
Pauli-Jordan function \cite{Pauli-Jordan} $\Delta
(x):=G_R(x)-G_A(x)$ and its analogue for a causal set. Note that
this could also generalize for causal sets embedable in curved
spacetimes.}.

\section{A Solution to the Cosmological Constant Problem}

The biggest probably success of the causal sets approach to date, in
terms of concrete result, is the prediction of the correct order of
magnitude of the cosmological constant \cite{Sorkin1990}. The
argument is heuristic, but conclusive, in the sense that a zero
cosmological constant (or of different magnitude) is ruled out by
causal sets (or else a different result rules out causal sets).
Another interesting point of this prediction is that it was first
made in 1990, when at the time it was believed that its value is
zero, and only much later was measured to be of that order of
magnitude. Here we will present briefly the argument, and the reader
is referred to \cite{Sorkin1990} where it first appeared and
\cite{everpresent} where it is analyzed in some depth along with
possible problems and open questions.

The cosmological constant problem arises from the following
observation. The dark energy is the 70\% of effective energy density
of the universe. Moreover it has negative pressure. This is usually
explained by the use of a cosmological constant $\Lambda$. However,
it appears to be very small, since the natural value we would expect
from the vacuum expectation value is $\rho_{vac}\simeq m_{pl}^4$
which is $\simeq 10^{120}\rho_{obs}$ the observed value. Most
mechanism that makes it small brings it to identically zero (e.g.
supersymmetry). The other strange feature is that the cosmological
constant is of the order of magnitude of the ambient density, but
only \emph{now} (i.e. in our epoch). This also seems as an unnatural
fine-tuning. The causal set proposal will suggest why the latter
fine tuning appears and explain how the cosmological constant is
non-zero (and of the desired magnitude) given a mechanism that would
otherwise bring it to zero.

We need to introduce here an alternative formulation of general
relativity the so-called \emph{unimodular} gravity. It is a path
integral formulation where the total volume $V$ is kept fixed. The
equation of motions are identically the same as in ordinary GR. The
cosmological constant term $\Lambda=\Lambda_0+\lambda$ appears in
the action along with $V$ in a $-\Lambda V$
term:

\beq\delta(\int(\frac1{2\kappa}R-\Lambda_0)dV-\lambda V)=0\eeq In
unimodular gravity, the cosmological constant and the volume are
conjugate pairs very much like Energy-Time. It obeys a similar
uncertainty principle where $\Delta V\cdot\Delta \Lambda\sim\hbar$.
In standard unimodular gravity, the volume is fixed, and the
cosmological constant is completely undetermined from the parameters
of the theory as expected.

However, in causal set things change. The role of time is replaced
by the number of elements $N$ (e.g. growth dynamics) and in QT we do
not sum over different times. It is natural to causal sets to
consider fixed $N$, which is \emph{almost} unimodular gravity. The
difference being that we have fixed $N$ instead of $V$. $N$ is
proportional to $V$ \emph{up to poisson fluctuations}, i.e. $\pm
\sqrt V$. We thus have some ``kinematic fluctuations''. Taking the
conjugate of the volume we get $\Delta \Lambda\sim1/\Delta
V\sim1/\sqrt V$. Assuming that there is a mechanism that brings the
cosmological constant to zero, then applying this on a causal set we
get $\Lambda\sim V^{-1/2}\sim H^2\sim \rho_{\textrm{critical}}$.
This cosmological constant is thus of the correct magnitude and is
\emph{always} of the order of magnitude of the critical density.
This happens by simple considerations of the kinematics of the
causal set (being a random partial order). The $V$ is the volume of
the past and thus result in an everpresent $\Lambda$ of varying
magnitude and sign. Particular models/implimantations of dynamics
are needed to check whether this assumption change the standard
picture of cosmology (as e.g. the Big Bang Nucleosynthesis). In
\cite{everpresent} such a model is considered.

\section{Summary and Conclusion}

We have briefly reviewed the causal sets approach to quantum
gravity, giving the original references for deeper study. In
particular we have seen what a causal set is and how it succeeds in
being both discrete and Lorentz invariant. The continuum concept
arise from solely the partial (causal) order, and it has been shown
that these do agree with the continuum notions for large enough
causal set that have a continuum spacetime that faithfully embeds in
it. Further we have seen how with some very simple and natural
requirements for the (classical) dynamics we get a family of
solutions that is quite restrictive. Getting the full quantum
dynamics of causal sets, is the major open issue for causal sets,
even though there are considerable recent developments. Firstly a
2-dimensional full gravity model that gives as the typical causal
set  Minkowski spacetime (which is certainly manifold-like!).
Secondly, transforming the Einstein action in terms of causal
relations has been done and it remains to be examined. Thirdly,
there is progress in the conceptual part, and the development of a
histories quantum theory of closed systems, the Piombino
interpretation. A semi-classical description, where spacetime is a
single causal set, but matter (and or fields) lying on it are
quantum has been developed recently in a satisfactory degree and
what remains now, is to explore possible deviations from standard
QFT on curved background physics, in order to test the assumption of
causal sets. Finally, we reviewed the proposed explanation, using
causal sets, for the small non-zero cosmological constant observed.

\ack The author would like to thank all the people that he
interacted and discussed about causal sets and in particular Rafael
Sorkin, Fay Dowker, Sumati Surya, David Rideout and Yousef
Ghazi-Tabatabai. This work is partly supported by IKY. Finally, the
author would like to thank the organizers of the conference.

\end{document}